\documentclass[12pt]{article}
\usepackage[utf8]{inputenc}
\usepackage[a4paper, top=2.5cm, bottom=2.5cm, left=2.2cm, right=2.2cm]
{geometry}
\usepackage{amsmath,amssymb,amsfonts,bm,amscd,mathtools}
\usepackage{graphicx}
\usepackage{caption, subcaption}
\usepackage{comment}
\usepackage{cite}
\usepackage{float}
\usepackage{appendix}
\usepackage{physics}
\usepackage{tikz}
\usepackage{tikz-cd}
\usetikzlibrary{cd}
\usepackage{array}
\usepackage[hidelinks]{hyperref}

\usepackage{tocbasic}
\DeclareTOCStyleEntry[
  beforeskip=.4em plus 1pt,% default is 1em plus 1pt
  pagenumberformat=\textbf
]{tocline}{section}

\newtheorem{example}{Example}

\newcommand{\Z}{{\mathbb Z}}

\newcommand{\C}{{\mathbb C}}
\newcommand{\Q}{{\mathbb Q}}

\def\be{\begin{equation}}
\def\ee{\end{equation}}

\title{Decorated TQFTs and their Hilbert Spaces}
\author{Mrunmay Jagadale}
\date{}

\begin{document}

\maketitle

\vspace{-1.5cm}

\begin{center}
    \textit{Walter Burke Institute for Theoretical Physics,  California Institute of Technology, Pasadena, CA 91125, USA}
\end{center}

\vspace{0.2cm}

\centerline{\small E-mail: mjagadal@caltech.edu}

\vspace{1cm}

\centerline{\bf Abstract} \bigskip
We discuss topological quantum field theories that compute topological invariants which depend on additional structures (or decorations) on three-manifolds. The $q$-series invariant $\hat{Z}(q)$ proposed by Gukov, Pei, Putrov and Vafa is an example of such an invariant. We describe how to obtain these decorated invariants by cutting and gluing, and make a proposal for Hilbert spaces that are assigned to two-dimensional surfaces in the $\hat{Z}$-TQFT.

\tableofcontents

\newpage

\section{Introduction}

When given a topological quantum field theory (TQFT), the first question one asks is, ``What does it compute?'' In general, given a three-manifold, a three-dimensional TQFT computes for us a topological invariant of that three-manifold. For example, the $SU(2)$ Chern-Simons theory at level $k\in \Z$ computes the Witten-Reshetikhin-Turaev (${\mathrm{WRT}}$) invariants of three-manifolds\cite{Witten:1988hf,Reshetikhin1991InvariantsO3}. A decorated TQFT computes a topological invariant that depends on additional data. We call this additional data ``decoration''.  One classic example of such an invariant is the Reidemeister-Milnor-Turaev torsion which is a topological invariant of three-manifolds that depends on the $\mathrm{Spin}^{c}$ structure of the three-manifold\cite{Turaev1997TorsionIO}. 

In \cite{Gukov:2016gkn, Gukov:2017kmk}, Gukov, Pei, Putrov and Vafa conjectured a three-manifold invariant $\hat{Z}_{b}(M_{3}, q)$ valued in $q$-series. This $q$-series invariant depends on the choice of $\mathrm{Spin}^{c}$ structure on the three-manifold. $\hat{Z}_{b}(M_{3},q)$ is believed to give a non-perturbative definition of complex Chern-Simons theory with gauge group $SL(2,\C)$. In various limits, this $q$-series invariant is related to other topological invariants \cite{Kucharski:2019fgh, Costantino:2021yfd, Chae:2022iuu}. It is connected to different areas of mathematics and physics such as resurgence\cite{Gukov:2016njj}, three-dimensional gauge theories, modular forms, vertex operator algebra \cite{Cheng20193dM, Bringmann:2018ddv, Cheng:2022rqr}, etc. 

In \cite{Atiyah1988TopologicalQF} Atiyah axiomatized the notion of quantum field theory. In a three-dimensional TQFT, a vector space is assigned to every two-dimensional surface $\Sigma$, and a vector in that vector space is assigned to a three-manifold with boundary $\Sigma$. We can obtain the partition function of a TQFT on a closed manifold by cutting it into simpler pieces and gluing them back together. Thus by their very nature TQFTs give us topological invariants. This axiomatization was extended to $\mathrm{Spin}$ TQFTs in \cite{Blanchet1995TopologicalAF, Blanchet1996TopologicalQF}. In this paper, we describe how to do the cutting and gluing for some TQFTs decorated by $\mathrm{Spin}^{c}$-structures or cohomology groups. 

Finding a four-dimensional TQFT that is a categorification of $\hat{Z}$-TQFT would be quite helpful for the classification problem of smooth four-manifolds. An important question for the categorification of $\hat{Z}$-TQFT is: What does $\hat{Z}$-TQFT assign to a circle? Or what is the category of line operators in $\hat{Z}$-TQFT? These questions are closely related to the problem of finding the Hilbert space associated to torus in $\hat{Z}$-TQFT.

In \cite{Gukov:2019mnk} Gukov and Manolescu introduced a two-variable series $F_{K}$ associated to three-manifolds with torus boundaries which can be thought of as a vector in Hilbert space associated to torus in $\hat{Z}$-TQFT. They also gave a formula for gluing them along the torus boundaries. In this paper, we explain how to express the cutting and gluing in terms of cutting and gluing of states and operators ($k$-linear maps) on Hilbert spaces and make a conjecture about the structure of Hilbert spaces in $\hat{Z}$-TQFT. We claim that the Hilbert space associated to genus $g$ surface $\Sigma_{g}$ in the $\hat{Z}$-TQFT is given by, 
\be 
\mathcal{H}_{\hat{Z}} (\Sigma_{g}) = \mathcal{H}_{(2g,2g)} \otimes \C[\Z^{2g}\times \Z^{2g}].
\ee 
Where $\mathcal{H}_{(2g,2g)}$ is the Hilbert space of $2g$ bosonic oscillators and $2g$ fermionic oscillators.

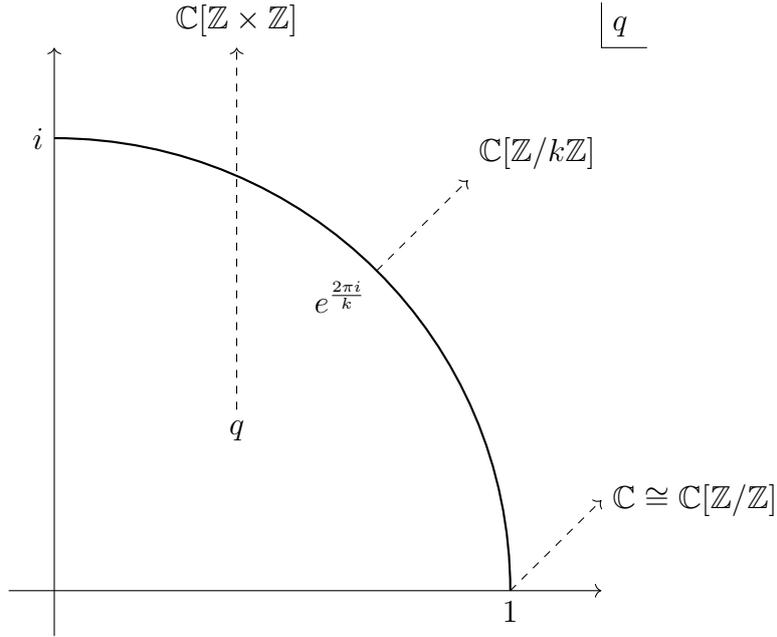
\begin{figure}
    \centering
\begin{tikzpicture}[scale=6]
   \draw[->] (-0.1cm,0cm) -- (1.2cm,0cm)  ;
   \draw[->] (0cm,-0.1cm) -- (0cm,1.2cm)  ;
   \draw[-] (1.3cm,1.2cm)-- (1.2cm,1.2cm) -- (1.2cm,1.3cm)  ;
   \node at (1.2cm,1.2cm) [above right] {$q$};
   \draw[thick] (0cm,0cm) (0:1cm)arc (0:90:1cm);
   \node at (1cm,0cm) [below] {$1$};
   \node at (0cm,1cm) [left] {$i$};
   \node at (0.707107cm,0.707107cm) [below left] {$e^{\frac{2 \pi i}{k}}$};
   \node at (0.4cm,0.4cm) [below] {$q$};
   \draw[->][dashed] (1cm,0cm) -- (1.2cm,0.2cm) ;
   \draw[->][dashed] (0.707107cm,0.707107cm) -- (0.907107cm,0.907107cm) ;
   \draw[->][dashed] (0.4cm,0.4cm) -- (0.4cm,1.2cm) ;
   \node at (1.2cm,0.2cm) [right] {$\C\cong \C[\Z/\Z]$};
   \node at (0.907107cm,0.907107cm) [above right] {$\C[\Z/k\Z]$};
   \node at  (0.4cm,1.2cm) [above] {$\C[\Z \times \Z]$};
\end{tikzpicture}
    \caption{0-decorated and $(\ell,m)$-graded sectors of Hilbert space associated to torus at $q=1$, $q=e^{\frac{2 \pi i}{k}}$ and at a generic value of $q$.}
    \label{0decoratedhilbertspaces}
\end{figure}

\noindent\textbf{Organization of the paper:} In section \ref{generalstructre}, we give a simple example of a decorated TQFT, where the TQFT is decorated by $H^{1}(M_{3},\Z_{2})$. We also discuss how the decorations of a three-manifold decompose into grading and decorations of Hilbert spaces in a decorated TQFT. In section \ref{inverseturaev}, we move on to a slightly non-trivial example of a decorated TQFT. We discuss the TQFT for inverse Reidemeister-Milnor-Turaev torsion which is decorated by $\mathrm{Spin}^{c}$-structures. In section \ref{Zhatsection}, we discuss how to obtain the $q$-series $\hat{Z}$ by cutting and gluing states and $k$-linear maps on a Hilbert space. In section \ref{ZNrWRTT}, using conjectured relations between the $\hat{Z}$-invariant and other three-manifold invariants we propose relations between Hilbert spaces in their TQFTs. These relations are illustrated in figure \ref{0decoratedhilbertspaces}. 

\section{General Structure of Decorated TQFTs}\label{generalstructre}

A simple example of decorated TQFT is $U(1)$ Chern-Simons theory at level $k$ ``enriched'' by $0$-form global symmetry $\Z_{2}$\footnote{In general, we could consider a $\C/\Z$ symmetry. However, here we consider it's subgroup $\Z_{2}\subset \C/\Z$. Where $(+1)\in \Z_{2} \rightarrow 0 \in \C/\Z $, and $(-1)\in \Z_{2} \rightarrow \frac{1}{2}\in \C/\Z$.}. We could think of this theory as Spin Chern-Simons theory, that was introduced in \cite{Dijkgraaf1990TopologicalGT}.  We couple this theory to a background flat connection $\frac{B}{2 \pi} \in H^{1}(M_{3},\Z_{2})$. Its partition function in terms of path integral can be written as, 
\be
Z(M_{3},B) = \int \mathcal{D}A \exp{ 2 \pi i  \int_{M_{3}} \frac{k A + B}{2 \pi} \wedge \frac{F}{2 \pi}}.
\ee 
The partition function depends on the topology of $M_{3}$ and additional data, viz. the background flat connection $B$. We say the partition function is decorated by $B$.

\begin{figure}[ht]
    \centering
    \begin{subfigure}{0.47\textwidth}
    \centering
    \includegraphics[scale=0.25]{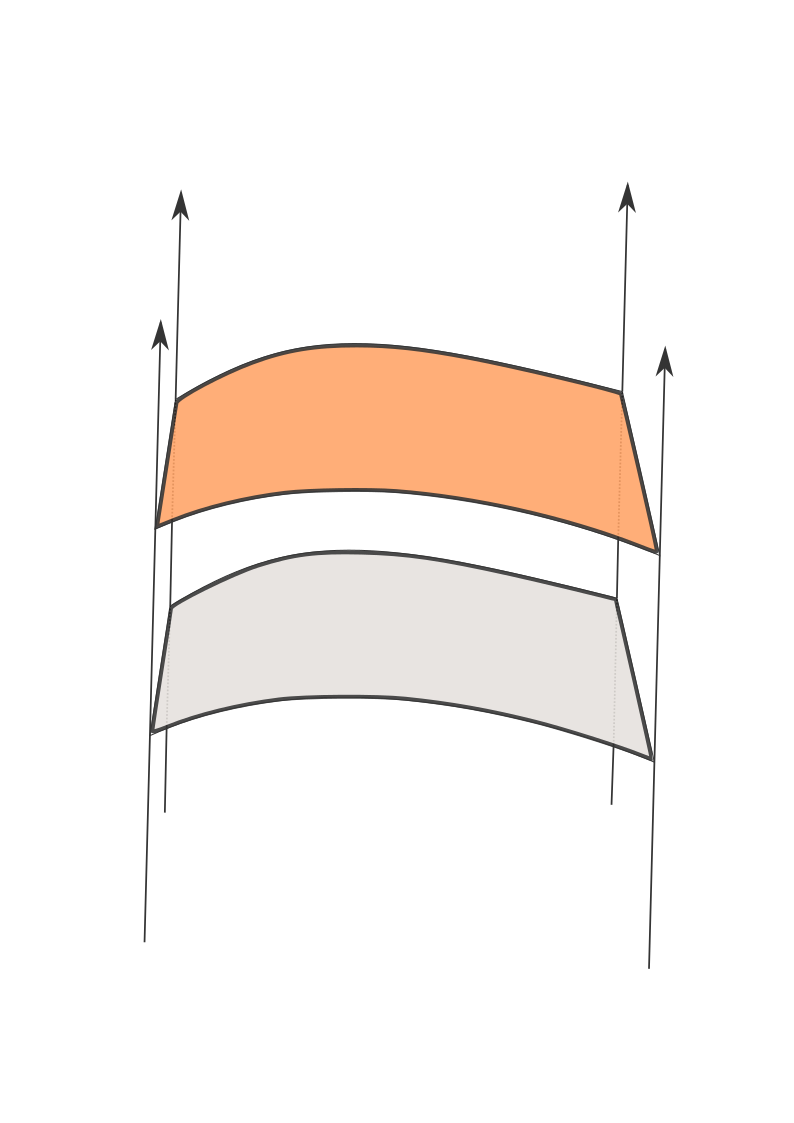}
    \caption{Charge operator on a ``constant time'' slice.}
    \label{chargeops}
    \end{subfigure}
    \begin{subfigure}{0.47\textwidth}
    \centering
    \includegraphics[scale=0.25]{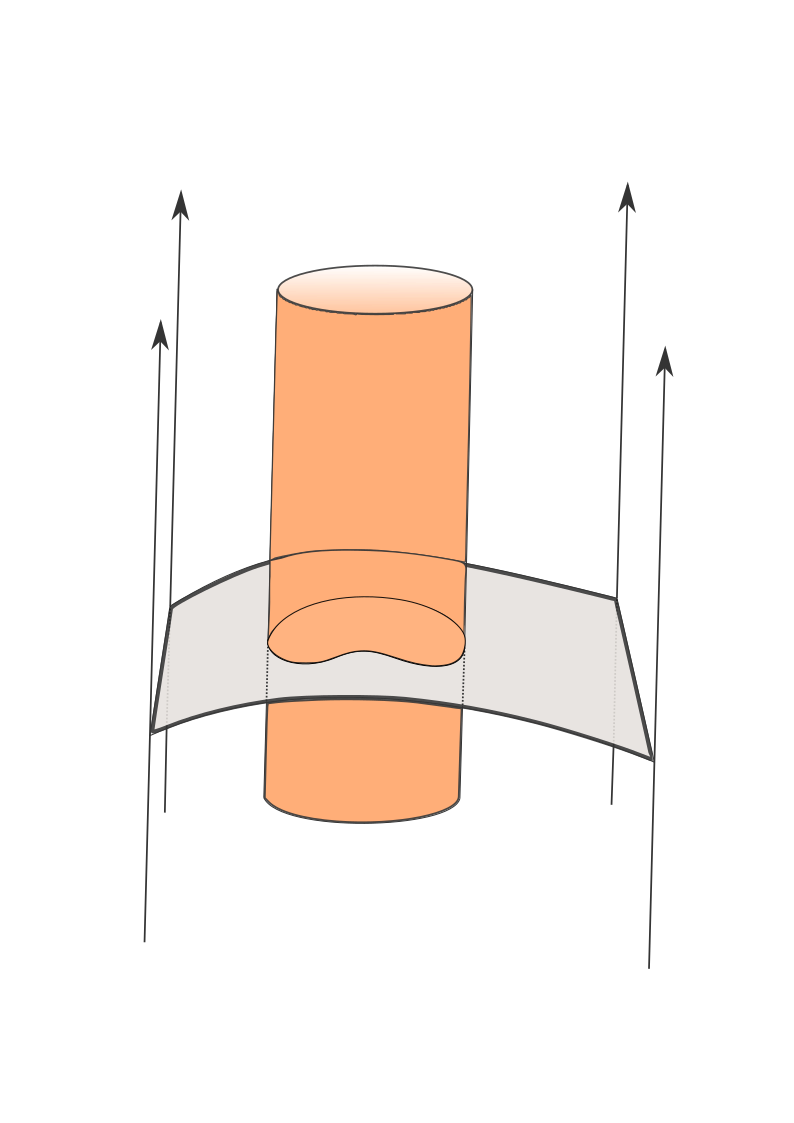}
    \caption{Charge operator with extent in ``time''.}
    \label{chargeopt}
    \end{subfigure}
    \caption{The arrows indicate the time direction, and the gray spatial slice is the slice with which we associated a Hilbert space.}
\end{figure}

This theory has $0$-dimensional charged operators, and charge operators $O_{g}(\Sigma)$ with $g\in \Z_{2} \subset  \C/\Z$ supported on a two dimensional surface $\Sigma$ (we refer to \cite{Gaiotto2014GeneralizedGS} for details on theories with generalised global symmetries). An operator with charge $m$ can be thought of as a monopole with magnetic flux $m= \int_{S^{2}} \frac{F}{2 \pi}$. The charge operator $O_{g}(\Sigma)$ is given by
\begin{align*}
    O_{g}(\Sigma)= \exp(i g \int_{\Sigma}F).
\end{align*}
Since $\int_{\Sigma}F= 2 \pi \Z$, these operators satisfy $O_{g_{1}}(\Sigma)\cdot O_{g_{2}}(\Sigma)=O_{g_{3}}(\Sigma)$ with $g_{3}= g_{1}+g_{2}\mod 1$. We can turn on the decoration $B$ by inserting a charge operator on $2$-chain representing the Poincaré dual of $B$. When it is inserted on a ``constant time'' slice (see figure \ref{chargeops}) we interpret it as an operator acting on Hilbert space, and when it has an extend in ``time-direction'' (see figure \ref{chargeopt}) it takes us to a different sector of the Hilbert space.
In general, Hilbert spaces associated with co-dimension 1 manifolds in a decorated TQFT have induced decorations and grading. A choice of decoration on $\Sigma \times S^{1}$ usually splits into a choice of decoration on $\Sigma$ and a choice of parameter dual to the grading on the Hilbert space associated with $\Sigma$. In our example, $H^{1}(\Sigma \times S^{1},\Z_{2})$ splits into,
\be
H^{1}(\Sigma \times S^{1},\Z_{2}) \cong H^{1}(\Sigma,\Z_{2}) \oplus \mathrm{Hom}(H_{0}(\Sigma,\Z), \Z_{2}).
\ee 
Thus we have a Hilbert space $\mathcal{H}(\Sigma)$ decorated by $H^{1}(\Sigma,\Z_{2})$ and graded by $H_{0}(\Sigma,\Z)$. The graded dimensions of this Hilbert space are given by 
\be 
Z(\Sigma\times S^{1}, \omega \oplus \alpha ) = \sum_{n\in H_{0}(\Sigma,\Z)} e^{2 \pi i \alpha(n)} \mathrm{dim}\mathcal{H}(\Sigma,\omega, n).
\ee 
Where $\omega \in H^{1}(\Sigma,\Z_{2})$ and $\alpha \in \mathrm{Hom}(H_{0}(\Sigma,\Z), \Z_{2})$, with $ \Z_{2}$ thought of as a subgroup of $\C/\Z$.

\begin{figure}[H]
    \centering
    \includegraphics[scale=1]{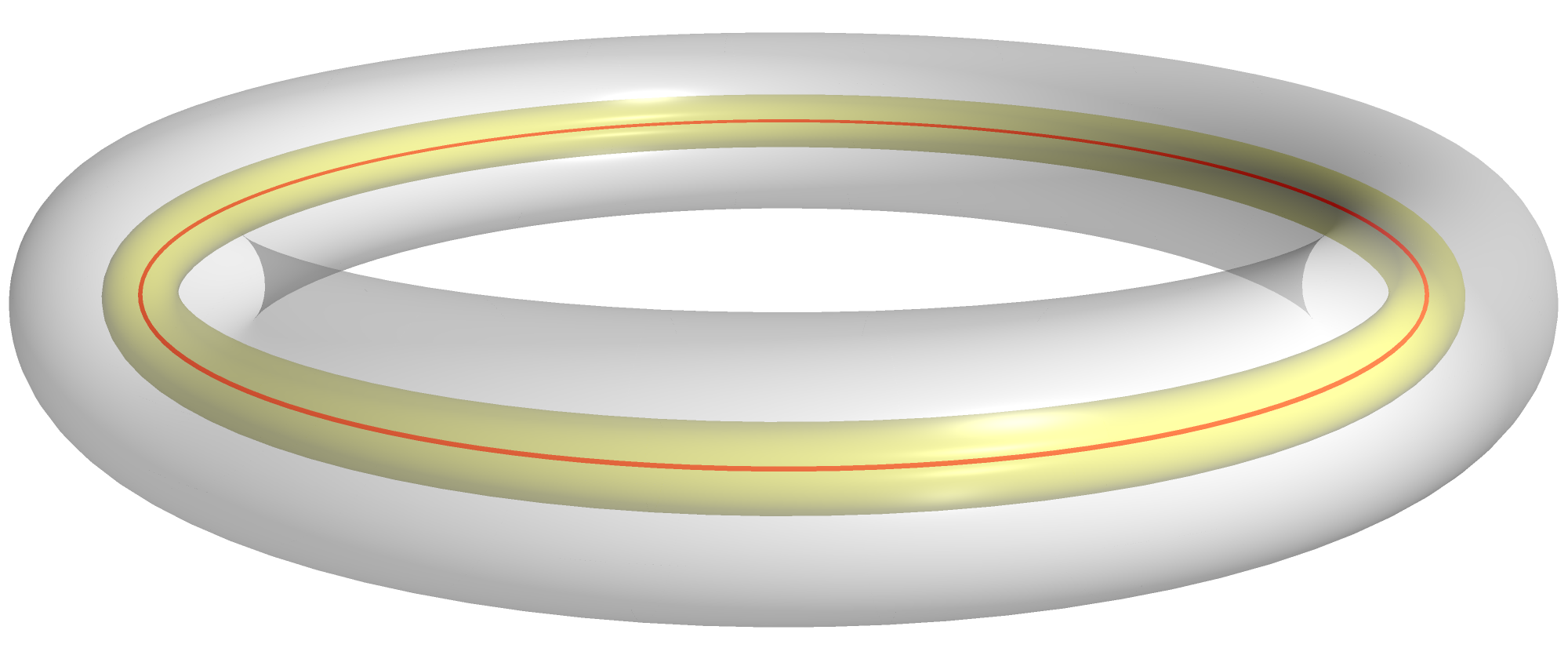}
    \caption{Line operator surrounded by a charge operator. }
    \label{dlineostorus}
\end{figure}

Another interesting set of operators in this theory is the set of line operators. The line operators are given by,
\be 
W_{e}(\gamma) = \exp(i e \int_{\gamma}A),
\ee 
where  the decimal part of $e$ is fixed, with $e= g \mod 1$. In $U(1)$ Chern-Simons theory at level $k$ ``enriched'' by $0$-form global symmetry $\Z_{2}$, there are $2k$ such line operators. These line operators have charges $[g],1+[g],\ldots,k-1+[g]$, where $[g]\in \{0,\frac{1}{2}\}$. Another way to think of these line operators is by thinking of the usual $U(1)$ Chern-Simons theory line operators sitting at the core of a solid torus with charge operator $O_{g}(\Sigma)$ surrounding them such that $\Sigma$ is homologous to the boundary torus (see figure \ref{dlineostorus}). Depending on how we fill in $T^{2}$ to get a solid torus, we get different vectors in the Hilbert space associated with the boundary torus.

For each decoration, the Hilbert space associated with torus is a $k$-dimensional vector space $\C^{k}$. The action of generators of modular group, $S$ and $T$, on these vector spaces is given by,
\begin{align}
    S_{\lambda_{1},\mu_{1},\lambda_{2},\mu_{2},n_{1},n_{2}}^{k_{1},k_{2}} &= \frac{1}{\sqrt{k}}\delta_{n1,n2} \delta_{\lambda_{1},-\mu_{2}} \delta_{\mu_{1},\lambda_{2}} q^{-(k_{1}+\lambda_{1})(k_{2}+\lambda_{2})}\\
    T_{\lambda_{1},\mu_{1},\lambda_{2},\mu_{2},n_{1},n_{2}}^{k_{1},k_{2}} &=\delta_{n1,n2}\delta_{\lambda_{1},\lambda_{2}}\delta_{\mu_{1},\mu_{2}-\lambda_{2}}\delta_{k_{1},k_{2}} e^{-\frac{i\pi }{12}} q^{\frac{(\lambda_{1}+k_{1})^{2}}{2}}.
\end{align} 
Here $\lambda_{i},\mu_{i} \in \C/\Z $, $n_{i}\in \Z $ give us the decorations, and $k_{i}\in \Z / k \Z$ label the basis of $\C^{k}$. In this example, the partition function is decorated by a flat connection $B\in H^{1}(M_{3},\Z_{2})$. Since the cohomology group $H^{1}(M_{3},\Z_{2})$ acts transitively and freely on $\mathrm{Spin}(M_{3})$, space of spin structures on $M_{3}$, we could think of $Z(M_{3},B)$ as being decorated by $\mathrm{Spin}(M_{3})$.

In general, the action of modular group on decorations and grading on $\Sigma$ tells us how different sectors labeled by decorations and grading are mapped to each other under the action of modular group on the Hilbert space. However, this does not completely specify the action of modular group on the Hilbert space. If the sectors of the Hilbert space with given decoration and grading are non-trivial, they could have a non-trivial action of the modular group.

\section{Inverse Reidemeister-Milnor-Turaev Torsion}\label{inverseturaev}
We will now look at topological quantum field theories decorated with $\mathrm{Spin}^{c}$-structures. Due to bijection between sets $\mathrm{Spin}^{c}(M_{3})$ and $H^{2}(M_{3},\Z) $, we can think of them as TQFTs decorated by $H^{2}(M_{3},\Z)$. Reidemeister-Milnor-Turaev torsion, $\tau$, is a $\mathrm{Spin}^{c}$ decorated topological invariant which can be computed by $U(1,1)$ supergroup Chern-Simons theory coupled to a background complex flat connection \cite{Rozansky:1992zt}. It is closely related to Alexander polynomial, whose TQFT construction was discussed in \cite{FroNic92}. Inverse Reidemeister-Milnor-Turaev torsion is a Laurent series in generators of the first homology group. By inverse Reidemeister-Milnor-Turaev torsion we mean the Laurent series we get by inverting Reidemeister-Milnor-Turaev torsion.

For example, the Reidemeister-Milnor-Turaev torsion for a mapping tori of $T^{2}$ is given by,
\be 
\tau(T^{2}\times_{\varphi}S^{1}) = \frac{\det(z I_{2\times 2} - \varphi)}{(z-1)^{2}} = \sum_{n\in\Z}  \delta_{n,0} +\frac{|n|}{2} (2 - \Tr \varphi) z^{n}  .
\ee 
Where $\varphi$ is the element of the mapping class group of torus (i.e. $\varphi \in SL(2,\Z)$) describing the twist along the base circle $S^{1}$, and $z$ is the generator of the cycle along the base circle. The inverse Reidemeister-Milnor-Turaev torsion for a mapping tori of $T^{2}$ is given by,
\begin{align}
 \tau^{-1}(T^{2}\times_{\varphi}S^{1})  =& \hspace{0.15cm} \frac{(z-1)^{2}}{\det(z I_{2\times 2} - \varphi)}.
\end{align}
Its Laurent series is given by,
\begin{align}
    &\sum_{n\in\Z}  \delta_{n,0} + \mathrm{sgn}(n)   \frac{ \Tr \varphi-2}{2} \frac{( \Tr \varphi+\sqrt{( \Tr \varphi)^{2}-4})^{n}-( \Tr \varphi-\sqrt{( \Tr \varphi)^{2}-4})^{n}}{2^{n} \sqrt{( \Tr \varphi)^{2}-4}}  z^{n}.
\end{align}
For mapping tori $T^{2}\times_{\varphi}S^{1} $,
\be 
H_{1}(T^{2}\times_{\varphi}S^{1} ) \cong \mathrm{Coker}(\varphi-I) \oplus \Z.
\ee 
Using the bijection between the sets $\mathrm{Spin}^{c}(M_{3})$ and $H^{2}(M_{3},\Z)  \cong H_{1}(M_{3},\Z)$, we can get the $\mathrm{Spin}^{c}(M_{3})$ dependence of $\tau$ or its inverse. $\tau(T^{2}\times_{\varphi} S^{1})$ or its inverse doesn't depend on the generators in $\mathrm{Coker}(\varphi-I)$. In other words, they are non-zero only for $0\in \mathrm{Coker}(\varphi-I)$. For $n\in \Z$ $\tau$ and $\tau^{-1}$ are given by the coefficient of $z^{n}$ in their respective series.  

If we think of $ \tau^{-1}(T^{2}\times_{\varphi}S^{1})$ as a partition function of a quantum field theory, it suggest that the factor $(z-1)^{2}$ is coming from fermionic states, while $\det(z I_{2\times 2} - \varphi) $ is coming from bosonic states. As we will see, this is indeed the case. This TQFT is related to the TQFT that computes $\tau$ by sending the fermions that give the factor $(1-z)^{2}$ to bosons and sending the bosons that give the factor $\det(z I_{2\times 2} - \varphi) $ to fermions. 

The TQFT that computes the inverse Reidemeister-Milnor-Turaev torsion is decorated by $H^{2}(M_{3},\Z)$. $H^{2}(\Sigma\times S^{1},\Z)$ splits into,
\be 
H^{2}(\Sigma\times S^{1},\Z) \cong H^{2}(\Sigma,\Z)\oplus  H_{1}(\Sigma,\Z).
\ee 
Therefore, the Hilbert space associated to $\Sigma$ in this TQFT is decorated by $H^{2}(\Sigma,\Z)$ and graded by $\mathrm{Hom}(H_{1}(\Sigma,\Z),\C/\Z)\cong H^{1}(\Sigma,\C/\Z)$. Let's now look at the Hilbert space associated to torus in this TQFT.  It is given by,
\be 
\mathcal{H}_{\tau^{-1}} (T^{2}) =  \C[\C/\Z \times \C/\Z]\otimes \mathcal{H}_{(2,2)}. 
\ee 
Where $\mathcal{H}_{(2,2)}$ is the Hilbert space of two fermionic and two bosonic harmonic oscillators. The decoration $H^{2}(T^{2},\Z)\cong \Z$ is given by the particle number on $\mathcal{H}_{(2,2)}$. While the grading $\mathrm{Hom}(H_{1}(T^{2},\Z),\C/\Z) \cong \mathrm{Hom}(\Z^{2},\C/\Z) \cong (\C/\Z)^{2}$ is inherited from the $(\C/\Z)^{2}$ grading of $\C[\C/\Z \times \C/\Z]$. 

Let $\alpha$ and $\beta$ be the bosonic annihilation operators, and $\psi$ and $\chi$ be the fermionic annihilation operators in $\mathcal{H}_{(2,2)}$. Their non-trivial (anti)commutation relations are given as follows,
\begin{align}\label{commutationinvturaev}
    [\alpha,\alpha^{\dag}]&=z  &  [\beta,\beta^{\dag}]&=z &
    \{ \psi,\psi^{\dag} \} &=z &  \{ \chi,\chi^{\dag} \} &=z.
\end{align}
The basis of $n$-particle subspace of $\mathcal{H}_{(2,2)}$  consists of states of the form 
\begin{align}
\ket{i,n-a-b-i,a,b} &= \frac{1}{\sqrt{i! (n-a-b-i)!}} (\psi^{\dag})^{a}(\chi^{\dag})^{b}   (\alpha^{\dag})^{i}(\beta^{\dag})^{n-a-b-i} \ket{0}
\end{align}
Where $a,b\in\{0,1\}$, and $i,n-a-b-i\in\{0,1,\ldots n-a-b\}$.
With (anti)commutation relations given in equation \eqref{commutationinvturaev}, the norms of $n$-particle states described above are simply $(-1)^{a+b}z^{n}$. 

The Hilbert space $\mathcal{H}_{(2,2)}$ can be broken down into four subspaces; one purely bosonic, two with one fermionic particle and one with two fermionic particles.  Further, the vector space of purely bosonic states  can be written as direct sum of symmetric tensor products of purely bosonic one particle subspace.
\begin{align}
\mathcal{H}_{(2,2)}&=\left[ \bigoplus_{n =0}^{\infty} \mathrm{Sym}^{n} V \right]\oplus \psi^{\dag} \left[ \bigoplus_{n =0}^{\infty} \mathrm{Sym}^{n} V \right]\oplus \chi^{\dag}\left[ \bigoplus_{n =0}^{\infty} \mathrm{Sym}^{n} V \right]\oplus\psi^{\dag} \chi^{\dag}\left[ \bigoplus_{n =0}^{\infty} \mathrm{Sym}^{n} V \right].
\end{align}
Where $V$ is the two dimensional vector space $V=\mathrm{Span}\{\alpha^{\dag}\ket{0},\beta^{\dag}\ket{0}\}$. This division into four subspaces carries on to the $n$-particle subspace of $\mathcal{H}_{(2,2)}$. For mapping tori of torus, the part of partition function coming from fermions, $(z-1)^{2}$, does not dependent on twisting along base circle. This tells us that the action of $SL(2,\Z)$ on fermionic generators is trivial. Therefore, the action $\varphi \in SL(2,\Z)$ on $n$-particle subspace of $\mathcal{H}_{(2,2)}$ takes the following block diagonal form,
\begin{align}
    \mqty(\dmat{\varphi_{n},\varphi_{n-1},\varphi_{n-1},\varphi_{n-2}}).
\end{align}
Where $\varphi_{n}$ represents the action of $\varphi$ on purely bosonic $n$-particle subspace $\mathrm{Sym}^{n} V $. 
The action of $\varphi$ on $\mathrm{Sym}^{n} V $ given by its action on $V$, which is the usual action of $SL(2,\Z)$ on a two dimensional vector space. $\mathrm{Sym}^{n} V $ is a $n+1$-dimensional vector space with basis 
\begin{equation}
   \left\{ \ket{i} \Big\vert \ket{i}=\frac{1}{\sqrt{i! (n-i)!}} (\alpha^{\dag})^{i}(\beta^{\dag})^{n-i} \ket{0}, i \in \{0,1,\ldots,n\}\right\}.
\end{equation}
In this basis the matrix elements of $\varphi = \begin{pmatrix} a &b \\ c& d \end{pmatrix} \in SL(2,\Z)$ are given by,
\begin{align}
    \varphi_{i,j} &= \sum_{k=\mathrm{Max}(i,j)}^{\mathrm{Min}(i+j,n)} \frac{a^{n-k}b^{k-i}c^{k-j}d^{i+j-k}}{(k-i)!(k-j)!(n-k)!(i+j-k)!} \sqrt{i ! (n-i)! j! (n-j)! }.
\end{align}

Now let's look at the $SL(2,\Z)$ action on $\C[\C/\Z \times \C/\Z]$ part of the Hilbert space. We consider the basis of $\C[\C/\Z \times \C/\Z]$ labeled by $(\lambda,\mu)\in(\C/\Z)^{2}$, $\{f_{\lambda,\mu}| f_{\lambda,\mu}(\lambda^{\prime},\mu^{\prime}) = \delta(\lambda-\lambda^{\prime})\delta(\mu-\mu^{\prime})\}$. In this basis the matrix elements of $\varphi = \begin{pmatrix} a &b \\ c& d \end{pmatrix} \in SL(2,\Z)$ are given by\footnote{Note the delta function is on $\C/\Z$.},
\be 
\varphi_{\lambda_{1},\mu_{1},\lambda_{2},\mu_{2}} =\delta(a \lambda_{1}+c \mu_{1}-\lambda_{2} ) \delta(b \lambda_{1}+d \mu_{1}-\mu_{2}).
\ee  
Taking a graded trace of $\varphi:\mathcal{H}_{\tau^{-1}(T^{2})}\rightarrow\mathcal{H}_{\tau^{-1}(T^{2})}$ gives us the inverse Reidemeister-Milnor-Turaev torsion, $\tau^{-1}_{\ell,m}(T^{2}\times_{\varphi}S^{1},z)$, of mapping tori $T^{2}\times_{\varphi}S^{1}$. Taking a trace over $\mathcal{H}_{2,2}$ gives us,
\be 
\Tr_{\mathcal{H}_{2,2}} (\varphi) = \frac{(z-1)^{2}}{1-(a+d)z+z^{2}}   .  
\ee 
While, taking a graded trace of $\varphi$ over $\C[\C/\Z \times \C/\Z]$ gives us,
\begin{align}
  \int_{0}^{1}\int_{0}^{1} \dd \lambda \dd \mu \hspace{0.1cm} \varphi_{\lambda,\mu,\lambda,\mu} e^{2 \pi i (\lambda \ell + \mu m )}   & = \int_{0}^{1}\int_{0}^{1} \dd \lambda \dd \mu \hspace{0.1cm} \delta(a \lambda+c \mu-\lambda ) \delta(b \lambda+d \mu-\mu) e^{2 \pi i (\lambda \ell + \mu m )}  \nonumber \\ 
  & = \sum_{k_{\ell},k_{m} \in \Z} \int_{0}^{1}\int_{0}^{1} \dd \lambda \dd \mu\hspace{0.1cm}  e^{- 2 \pi i \left( k_{\ell}(a \lambda+ c \mu -\lambda)+k_{m} (b \lambda+d \mu-\mu) \right)}e^{2 \pi i (\lambda \ell + \mu m )} \nonumber \\
  &= \sum_{k_{\ell},k_{m} \in \Z} \delta_{\ell,(a-1) k_{\ell}+b k_{m}} \delta_{m,c k_{\ell}+(d-1)k_{m} }.
\end{align}
Note the graded trace of $\varphi$ over $\C[\C/\Z \times \C/\Z]$ is non-zero only for $(\ell,m)\in (\varphi-I)\Z^{2}$ that is $(\ell,m)=0 \in \mathrm{Coker}(\varphi-I)$.

We can represent the Hilbert space in such a way that it is graded by $H^{1}(T^{2},\Z)$ and decorated by $H^{2}(T^{2},\Z)$ instead of working with Pontryagin dual grading $H^{1}(T^{2},\Z)$. In that case, the Hilbert space associated with torus can be written as 
\be 
\mathcal{H}_{\tau^{-1}} (T^{2}) = \C[\Z^{2}]\otimes \mathcal{H}_{2,2}.
\ee
For genus $g$ surface $\Sigma_{g}$, Hilbert space associated with it in this TQFT is given by 
\be 
\mathcal{H}_{\tau^{-1}} (\Sigma_{g}) = \C[\Z^{2g}]\otimes \mathcal{H}_{2g,2g}.
\ee 
Where $\mathcal{H}_{2g,2g}$ is the Hilbert space of $2g$-bosonic oscillators and $2g$-fermionic oscillators. $Sp(2g,\Z)$ acts trivially on the fermionic creation operators. The action of $Sp(2g,\Z)$ on bosonic operators is induced by its action on the $2g$-dimensional vector space of one-particle bosonic states. The action of $Sp(2g,\Z)$ on $\C[\Z^{2g}]$ is induced by action of $Sp(2g,\Z)$ on $\Z^{2g}$.

\section{\texorpdfstring{$q$}{q}-series \texorpdfstring{$\hat{Z}$}{Z}}\label{Zhatsection}

Since the $q$-series invariant $\hat{Z}(q)$ was first proposed in \cite{Gukov:2016gkn,Gukov:2017kmk}, the understanding of its decorations has developed over time. In \cite{Gukov:2016gkn,Gukov:2017kmk} $\hat{Z}(q)$ was labeled by abelian flat connections. For rational homology spheres, the set of flat abelian connections is the same as $H_{1}(M_{3},\Z)/\Z_{2}$. In \cite{Chun:2019mal}, for manifolds with $b_{1}>0$, $\hat{Z}$ was decorated by abelian and ``almost abelian" flat connections on $M_{3}$. The set of abelian flat connections, in this case, is in bijection with the torsion part of $H_{1}(M_{3},\Z)/\Z_{2}$. Later, in \cite{Gukov:2019mnk} it was understood that $\hat{Z}$ should in fact, be decorated by $\mathrm{Spin}^{c}$-structures on $M_{3}$. Using the bijection between $\mathrm{Spin}^{c}(M_{3})$ and $H_{1}(M_{3},\Z)$, the $\hat{Z}$s labeled by abelian flat connections now correspond to $\hat{Z}$s labeled by spin$^{c}$ structures associated with $(0,b) \in H_{1}(M_{3},\Z)\cong \Z^{b_{1}} \oplus \mathrm{Tors}H_{1}(M_{3},\Z)$. Where $\mathrm{Tors}H_{1}(M_{3},\Z)$ is the torsion part of $H_{1}(M_{3},\Z)$.

In this section we will interpret the surgery formula for $\hat{Z}$ on plumbed manifolds proposed in \cite{Chun:2019mal} as cutting and gluing of states and operators ($k$-linear maps) on a Hilbert space assigned to a torus, and make comments on how this Hilbert space is related to the Hilbert space that $\hat{Z}$-TQFT assigns to a torus.

\subsection*{Surgery Formula for Plumbed Manifolds}
By the Lickorish–Wallace theorem any closed oriented connected 3-manifold can be obtained by performing an integral Dehn surgery on a link in $S^{3}$. Plumbed manifolds are special class of manifolds which can be obtained by performing an integral Dehn surgery on a link in $S^{3}$ which is made up of linked unknots. This class of three-manifolds can be described by a graph whose vertices are labeled by integers. This graph is called the plumbing graph. 

\begin{figure}[H]
    \centering
    \includegraphics[scale=0.3]{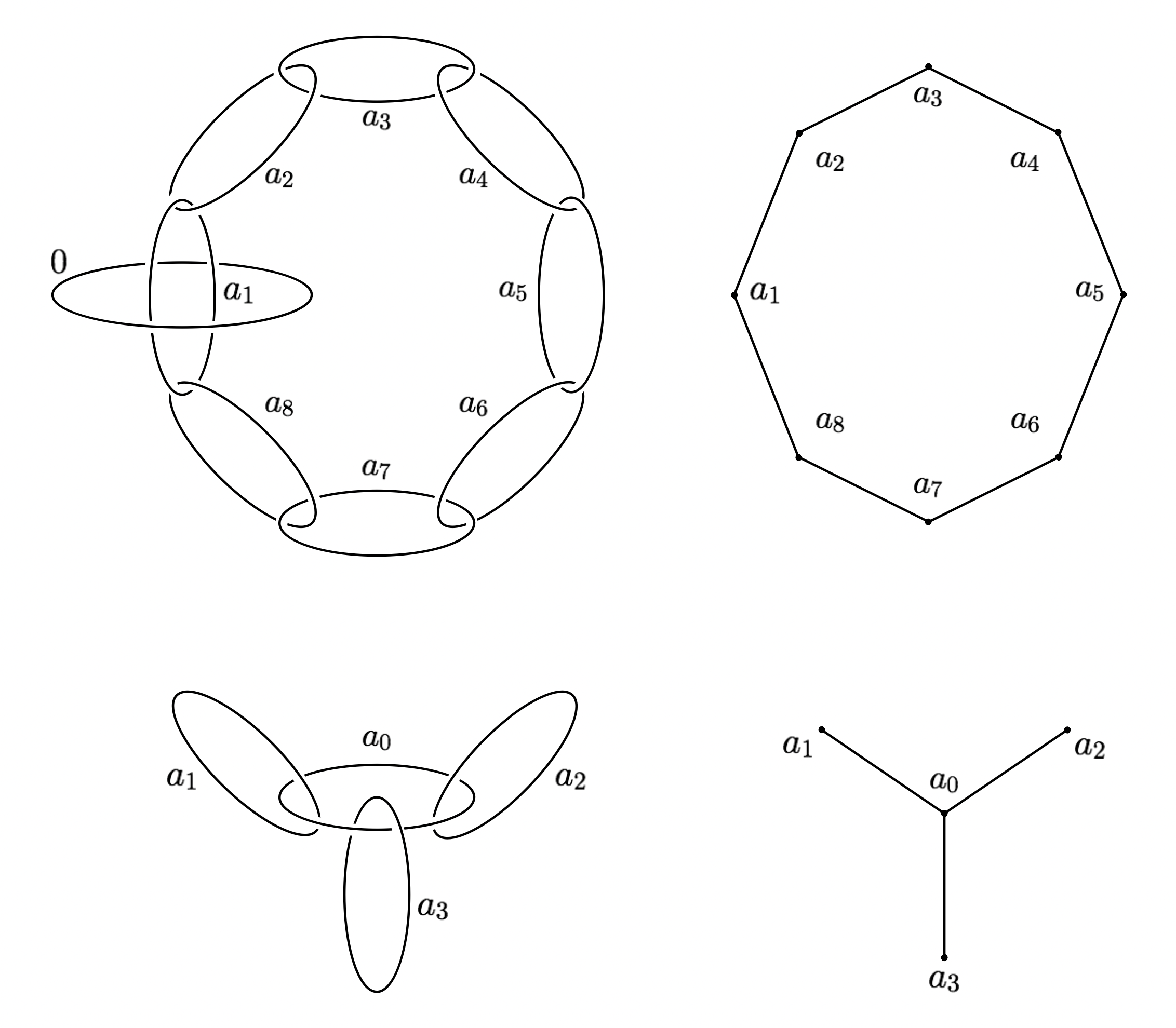}
    \caption{ Examples of plumbing graph of links of unknots.  }
    \label{plumbinglinkgraph}
\end{figure}

Each vertex of the plumbing graph corresponds to an unknot in $S^{3}$ and the integer that labels the vertex is the framing of that unknot. Edge between two vertices corresponds to a linking between the unknots corresponding to the two vertices. For each cycle in the plumbing graph we add a $0$-framed unknot that wraps around the cycle (see figure \ref{plumbinglinkgraph}).   

The plumbing graph can further be described by its linking matrix, which is defined as follows,
\begin{align}
    Q_{v v^{\prime}} &= \begin{cases}  a_{v} &\text{ if } v = v^{\prime} \\
                                       -1    &\text{ if } (v,v^{\prime})\in E \\ 
                                       0 & \text{otherwise.}
              \end{cases}
\end{align}
Where $v, v^{\prime}$ are in the vertex set of the plumbing graph, $E$ is the edge set, and $a_{v}$ are the framing coefficients. The first homology group of the plumbed manifold can be described in terms of its linking matrix as follows,
\be 
H_{1}(M_{3},\Z) = \Z^{b_{1}(\Gamma)} \times \Z^{V}/Q \Z^{V}.
\ee 
Where $V$ is the number of vertices in the plumbing graph, and $b_{1}(\Gamma)$ is the first Betti number of the graph, or equivalently number of cycles in the graph. 

In \cite{Chun:2019mal} a surgery formula for $\hat{Z}$ of plumbed manifolds with $b_{1}>0$ was given. This surgery formula gives us $\hat{Z}_{0,b}$, with $(0,b)\in \Z^{b_{1}} \oplus (2\mathrm{Coker}(Q)+\delta)/\Z_{2}$. The surgery formula for $\hat{Z}_{0,b}$ can be written as 
\begin{align}\label{plumbing1}
\hat{Z}_{0,b}(q) =&  q^{\frac{3 \sigma - \sum_{v}a_{v}}{4}} q^{-\frac{b^{T} Q^{-1} b}{4}}  \sum_{k \in \Z^{V}}  q^{- ( k^{T} Q k + k^{T} b )} \textrm{v.p.} \oint_{|z_{v}|=1} \prod_{v} \frac{\dd z_{v}}{2 \pi i z_{v}} \left( z_{v} -\frac{1}{z_{v}} \right)^{2-\mathrm{deg}(v)} z^{2Q k + b }.
\end{align}
Where $\sigma$ is the signature of the linking matrix $Q$, $\mathrm{deg}(v)$ is the degree of vertex $v$, and ``$\mathrm{v.p.}$" tells us that we should consider principle value prescription for contour integrals (for more details we refer to \cite{Chun:2019mal}). Let $f_{Q,n_{v}}$ denote the coefficients of the series expansion of $(z_{v}-1/z_{v})^{2-\mathrm{deg}(v)}$. That is, 
\be 
( z_{v} -1/z_{v} )^{2-\mathrm{deg}(v)} = \sum_{n_{v} \in \Z} f_{Q,n_{v}} z^{-n_{v}}_{v}
\ee 
$f_{Q,n_{v}}$ is simple for $\mathrm{deg}(v)\leq 2$ and terminates after finite terms. For $\mathrm{deg}(v) > 2$,  $f_{Q,n_{v}}$ is given by, 
\begin{align}
    f_{Q,n_{v}} &= \begin{cases} 
                                 \frac{\mathrm{sgn}(n_{v})^{\mathrm{deg}(v)}}{2}  \binom{\frac{\mathrm{deg}(v)+|n_{v}|}{2}-2}{\mathrm{deg}(v)-3} &\text{ if } |n_{v}|\geq \deg(v)-2, \text{ and } n_{v} = \mathrm{deg}(v) \mod 2 \\ 
                                 0  & \text{otherwise.}   \end{cases}
\end{align}
Using the series expansion of $(z_{v}-1/z_{v})^{2-\mathrm{deg}(v)}$, we can do the principle value prescription contour integrals in equation \eqref{plumbing1} and get,
\be \label{plumbing2}
\hat{Z}_{0,b}(q) = q^{\frac{3 \sigma - \sum_{v}a_{v}}{4}} q^{-\ell k(\frac{b}{2},\frac{b}{2}) }\sum_{k \in \Z^{V}} \sum_{n \in \Z^{V}}   q^{- \chi_{b}(k)}   f_{Q,n}  \delta_{2Q k + b,n} 
\ee 
Where $\ell k: \mathrm{Tors}H_{1}(M_{3})\times \mathrm{Tors}H_{1}(M_{3})\rightarrow \Q/\Z$, is the linking pairing, which is given by $\ell k(a,b)= a^{T}Q^{-1}b \mod 1$, the quadratic function $\chi_{b}: \Z^{V} \rightarrow \Z$ is given by $\chi_{b}(k) = k^{T}Q k +  b^{T} k $, and the term $q^{\frac{3 \sigma - \sum_{v}a_{v}}{4}}$ comes from the framing anomaly. Since the quadratic function $\chi_{b}$ is valued in integers, the sum in equation \eqref{plumbing2} is valued in $\Z[[q,q^{-1}]]$\footnote{For negetive definite plumbed manifolds we can choose $b$ such that the sum is valued in $\Z[[q]]$.}.

\subsection*{Surgery Formula from Hilbert Space}

\begin{figure}[H]
    \centering
    \includegraphics[scale=0.2]{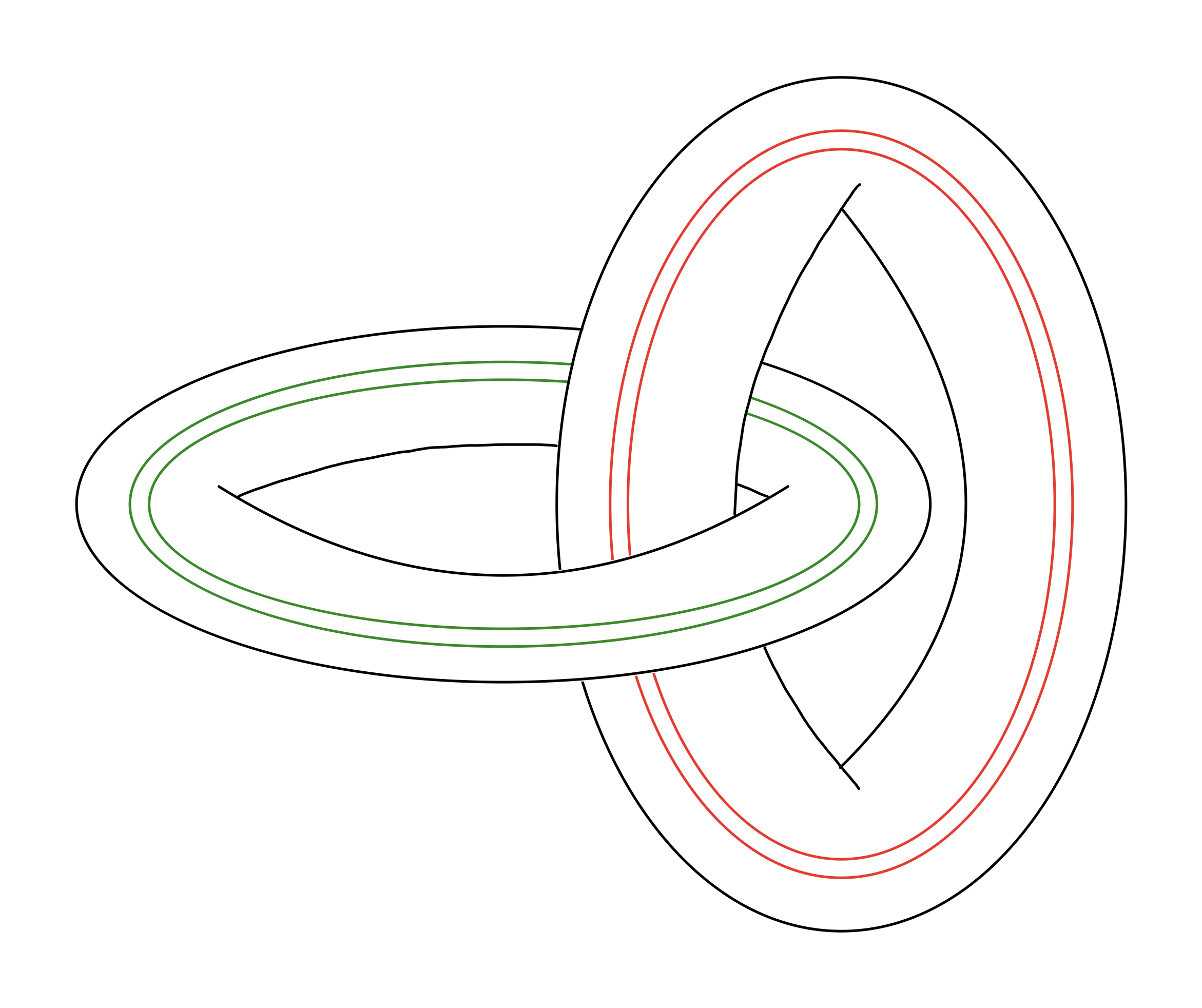}
    \caption{ Gluing two $T^{2}\times I$ along (black) boundary $T^{2}$ by $S\in SL(2,\Z)$.  }
    \label{gluingalongs}
\end{figure}

\begin{figure}[H]
    \centering
    \includegraphics[scale=0.22]{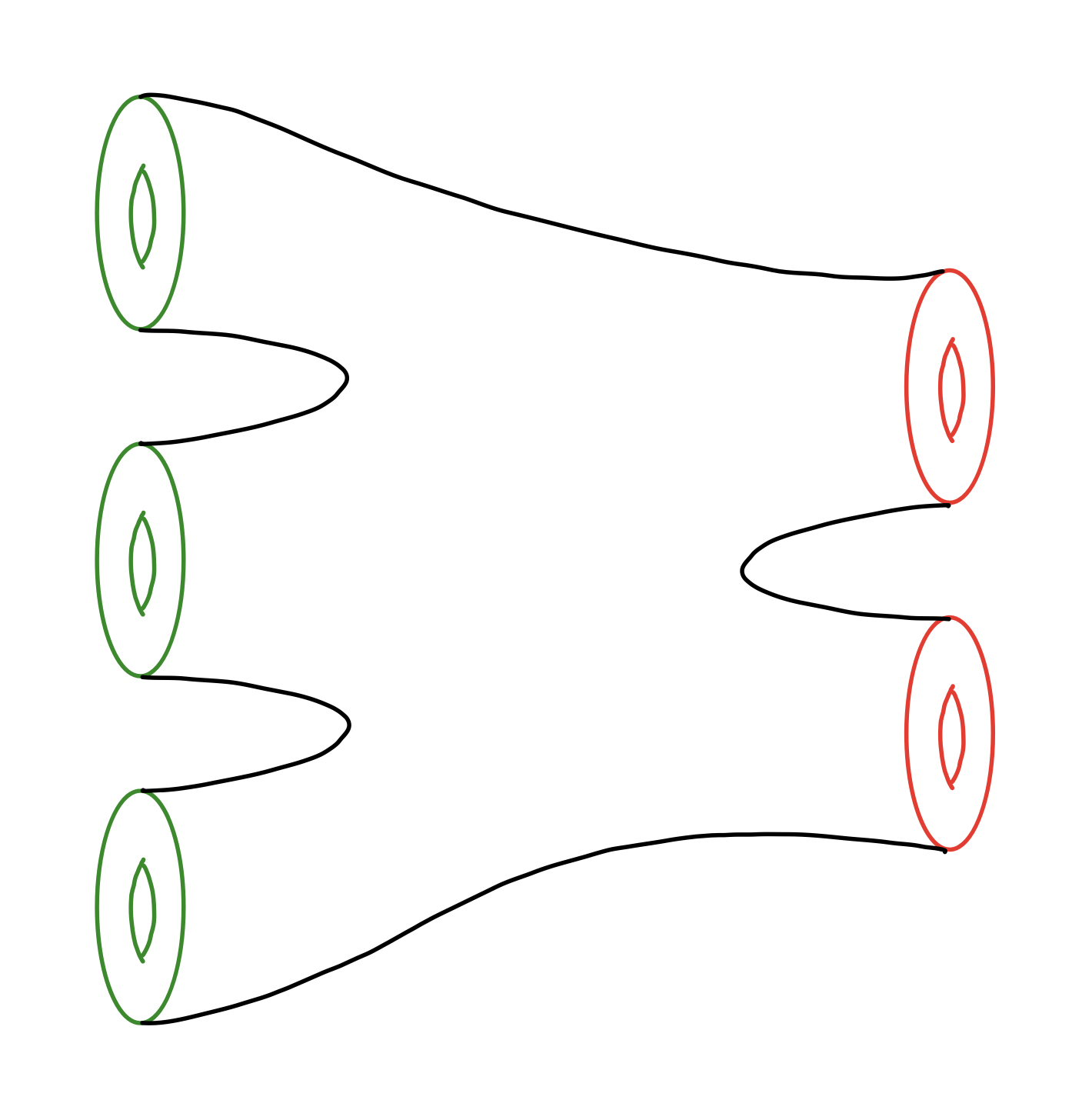}
    \caption{ This manifold is associated with an element of $\mathrm{Hom}(\mathcal{H}(T^{2})^{3},\mathcal{H}(T^{2})^{2})$. }
    \label{3mwtb}
\end{figure}

A plumbing graph of a plumbed three-manifold encodes the information about how the three-manifold can be obtained by gluing $T^{2}\times [0,1]$ along the torus boundaries. Each edge of plumbing graph corresponds to gluing by $S\in SL(2,\Z)$ (see figure \ref{gluingalongs}) and a vertex with coefficient $a_{v}$  corresponds to gluing by $T^{a_{v}}\in SL(2,\Z)$.
Similarly, a surgery formula encodes how a three-manifold invariant can be obtained by cutting and gluing. In a TQFT, a manifold with a torus boundary, depending on its orientation, is associated with a vector in $\mathcal{H}(T^{2})$ or $\mathrm{Hom}(\mathcal{H}(T^{2}),\C)$, a manifold with $r+r^{\prime}$ torus boundaries, with $r$ of them oriented one way and the other $r^{\prime}$ oriented the other way, is associated with an element of $\mathrm{Hom}(\mathcal{H}(T^{2})^{r},\mathcal{H}(T^{2})^{r^{\prime}})$(see figure \ref{3mwtb}). We want to understand how to get the surgery formula \eqref{plumbing2} by cutting and gluing states and operators ($k$-linear maps) on $\mathcal{H}(T^{2})$.  
To cut down $q^{-\frac{b^{T}}{2}Q^{-1}\frac{b}{2}} $ into pieces which can be glued, it is convenient to express it as $q^{-\beta^{T} Q \beta }$. Where $\beta$ is given by $\beta = \frac{1}{2}Q^{-1} b$. Note  $\beta \in (\Q/\Z)^{V}$ since, under $b\rightarrow b+ 2 Q x$, with $x\in \Z^{V}$, $\beta \rightarrow \beta + x$. Now the surgery formula \eqref{plumbing2} can be written as 
\begin{align}\label{plumbing3}
 \hat{Z}_{0,b}(q) = \sum_{\beta \in (\Q/\Z)^{V} } \hat{Z}^{\Q/\Z}_{0,\beta}(q) \delta_{2Q\beta,b},
\end{align}
where
\begin{align}\label{plumbing4}
  \hat{Z}^{\Q/\Z}_{0,\beta} (q) &=  q^{\frac{3 \sigma - \sum_{v}a_{v}}{4}} \sum_{n \in \Z^{V}} \sum_{k \in \Z^{V} } q^{- (k+\beta)^{T}Q(k+\beta) } f_{Q,n}\delta_{2Q(k+\beta),n}.
\end{align}
Note $ \hat{Z}^{\Q/\Z}_{0,\beta} (q)$ is non-zero only for $\beta$ such that $Q\beta \in \Z^{V}$. Written this way the summand in equation \eqref{plumbing4} can be broken down as follows,
\begin{align}
q^{- (k+\beta)^{T}Q(k+\beta) } & = q^{-\sum_{v}a_{v}(k+\beta)_{v}^{2} }  q^{2\sum_{(v,w)\in E}  (k+\beta)_{v} (k+\beta)_{w}}\label{breakdownofZ1}  \\
\delta_{2Q(k+\beta),n} & = \prod_{v} \delta_{2a_{v}(k+\beta)_{v}-2 \sum_{(v,w)\in E} (k+\beta)_{w},n_{v}}.\label{breakdownofZ2}
\end{align}
This suggests that the Hilbert space $\mathcal{H}(T^{2})$ is given by, 
\be \label{hilbertspaceplumbed}
\mathcal{H}(T^{2}) = \C[(\Q/\Z \times \Z) \times (\Q/\Z \times \Z)]\cong \C[\Q \times \Q].
\ee 
The fractional part $(\Q/\Z\times \Q/\Z)$ correspond to the label $\beta$. Further the equations \eqref{breakdownofZ1}, \eqref{breakdownofZ2} tell us that the matrix elements of $S$ and $T$ elements of $SL(2,\Z)$ in the basis given by $\{f_{\lambda,\mu}|f_{\lambda,\mu}(\lambda^{\prime},\mu^{\prime}) = \delta_{\lambda,\lambda^{\prime}}\delta_{\mu,\mu^{\prime}} ,\lambda,\mu \in \Q \}$ are,
\begin{align}
    S_{\lambda_{1},\mu_{1},\lambda_{2},\mu_{2}} &= q^{-2\mu_{1}\mu_{2}} \delta_{\lambda_{1},-\mu_{2}}\delta_{\mu_{1},\lambda_{2}} \\
    T_{\lambda_{1},\mu_{1},\lambda_{2},\mu_{2}} &= q^{-\mu_{1}^{2}} \delta_{\lambda_{1},\lambda_{2}+\mu_{2}}\delta_{\mu_{1},\mu_{2}}.
\end{align}
 In this basis, the matrix elements of $\varphi = \begin{pmatrix} a &b \\ c& d \end{pmatrix} \in SL(2,\Z)$ are given by,
\be 
\varphi_{\lambda_{1},\mu_{1},\lambda_{2},\mu_{2}} = q^{- c \lambda_{1} \lambda_{2} - b \mu_{1}\mu_{2}} \delta_{\lambda_{1},a \lambda_{2}+b \mu_{2}} \delta_{\mu_{1},c \lambda_{2}+d \mu_{2}}.
\ee 
Taking a trace of $\varphi$ we get $\hat{Z}$ of mapping tori $T^{2}\times_{\varphi}S^{1}$. We get the label $(\ell,m)$, with $(\ell,m)\in 2\mathrm{Coker}(\varphi-I)$, by inserting the operator $D(\lambda,\mu)$ in the trace, where $(\lambda,\mu)= \frac{1}{2} (\varphi-I)^{-1}(\ell,m)$. The operator $D(\lambda,\mu)$ is given by,
\begin{align}
     D(\lambda,\mu)_{\lambda_{1},\mu_{1},\lambda_{2},\mu_{2}} &= \sum_{k_{\ell},k_{m}\in \Z} \delta_{\lambda_{1},\lambda_{2}} \delta_{\mu_{1},\mu_{2}} \delta_{\lambda_{1},k_{\ell}+\lambda} \delta_{\mu_{1},k_{m}+\mu}.
\end{align}
Now $\hat{Z}_{0}$ for the mapping tori $T^{2}\times_{\varphi}S^{1}$ is given by,
\begin{align}
    \hat{Z}_{0,(\ell,m)} (T^{2}\times_{\varphi}S^{1}) & = \sum_{\lambda,\mu \in \Q/\Z} q^{\frac{\mathcal{L}-\mathcal{R}}{4}} \Tr[ \mathcal{D}(\lambda,\mu) \varphi ] \delta_{2 (\varphi-I)(\lambda,\mu),(\ell,m)}
\end{align}
Where $q^{\frac{\mathcal{L}-\mathcal{R}}{4}}$ is a contribution from ``framing anomaly". If $|\tr{\varphi}|>2$, we can represent the conjugacy class of $\varphi$ by $\pm R^{r_{1}}L^{\ell_{1}} \ldots R^{r_{n}}L^{\ell_{n}}$ with, $R=\begin{pmatrix} 1 &1 \\ 0& 1 \end{pmatrix}$, $L=\begin{pmatrix} 1 &0 \\ 1& 1 \end{pmatrix}$, and $r_{i},\ell_{i},n\geq 1$, then $q^{\frac{\mathcal{L}-\mathcal{R}}{4}} = q^{\frac{\sum \ell_{i} -\sum r_{i}}{4}}$.

\begin{example}\label{RLeg}
Let's look at an example with $\varphi = R L $. There is no anomaly contribution for $\varphi = R L $, therefore $ \hat{Z}_{0,(\ell,m)} (T^{2}\times_{R L}S^{1}) $ is given by
\begin{align}
    \hat{Z}_{0,(\ell,m)} (T^{2}\times_{R L}S^{1})  = & \sum_{\substack{\lambda_{1},\mu_{1}\in \Q \\ \lambda_{2},\mu_{2}\in \Q\\k_{\ell},k_{m}\in \Z}}
     \delta_{\lambda_{1},\lambda_{2}} \delta_{\mu_{1},\mu_{2}} \delta_{\lambda_{1},k_{\ell}+\frac{m}{2}} \delta_{\mu_{1},k_{m}+\frac{\ell- m}{2}}  q^{-  \lambda_{2} \lambda_{1} -  \mu_{2}\mu_{1}} \delta_{\lambda_{2},2 \lambda_{1}+ \mu_{1}} \delta_{\mu_{2}, \lambda_{1}+ \mu_{1}} \nonumber \\
     = & \sum_{k_{\ell},k_{m}\in \Z}  \delta_{\ell,2(k_{\ell}+k_{m})} \delta_{m, 2 k_{\ell}} 
\end{align}
Thus, $ \hat{Z}_{0,(\ell,m)} (T^{2}\times_{R L}S^{1}) $ is non-zero only for $(\ell,m) \in 2 (\varphi-I)\Z^{2}$ or equivalently  for $(\ell,m) = 0\in 2 \mathrm{Coker}(\varphi-I)$.
\end{example}

The vacuum state in the Hilbert space corresponds to the leaves of plumbing graph (degree one vertex). For a degree one vertex $v$ $f_{Q,n_{v}}$ is given by\footnote{Here by $d$ in $f_{d,n}$ we denote the degree of the vertex. Recall $f_{Q,n_{v}}$ only depends on degree of vertex $v$. } 
\be 
f_{1,n} = \delta_{n,-1} - \delta_{n,1}
\ee 
Therefore the vacuum state is given by
\begin{align}
    v_{\lambda,\mu} &=\sum_{\mu^{\prime} \in \Q/\Z} \delta_{\mu,\mu^{\prime}}\left(\delta_{2 \lambda,1}-\delta_{2 \lambda,-1}\right) &    v^{\dag}_{\lambda,\mu} &=\sum_{\mu^{\prime} \in \Q/\Z} \delta_{\mu,\mu^{\prime}}\left(\delta_{2 \lambda,-1}-\delta_{2 \lambda,1}\right)  .
\end{align}
While taking conjugate we take $\lambda \rightarrow -\lambda$, which accounts for orientation reversal. A degree $d>2$ vertex of plumbing graph corresponds to an operator $\mathcal{O}(a) \in \mathrm{Hom}(\mathcal{H}(T^{2})^{r},\mathcal{H}(T^{2})^{r^{\prime}}) $, with $r+r^{\prime}=d$ and where $a$ denotes the framing coefficient of the vertex. The operator $\mathcal{O}(a)$ is given by, 
\begin{align}
    \mathcal{O}(a)_{\lambda_{1},\mu_{1},\ldots\lambda_{r},\mu_{r}}^{{\tilde{\lambda}_{1},\tilde{\mu}_{1},\ldots\tilde{\lambda}_{r^{\prime}},\tilde{\mu}_{r^{\prime}}}} &= \sum_{n\in \Z} f_{d,n} q^{-a \mu_{1}^{2}  }  \delta_{2a \mu_{1}+ 2\sum_{1}^{r}\lambda_{i}+2\sum_{1}^{r^{\prime}}\tilde{\lambda}_{i},n} \prod_{i=2}^{r}\delta_{\mu_{1},\mu_{i}} \prod_{i=1}^{r^{\prime}}\delta_{\mu_{1},-\tilde{\mu}_{i}}
\end{align}

\begin{example}\label{3pteg} 
Lets look at an example where the plumbed manifold given by the plumbing graph from figure \ref{3ptegdiag}. The second cohomology group of this plumbed manifold is $\Z_{3}$.
\begin{figure}[H]
\centering
    \begin{subfigure}{0.45\textwidth}
    \centering
    \includegraphics[scale=0.25]{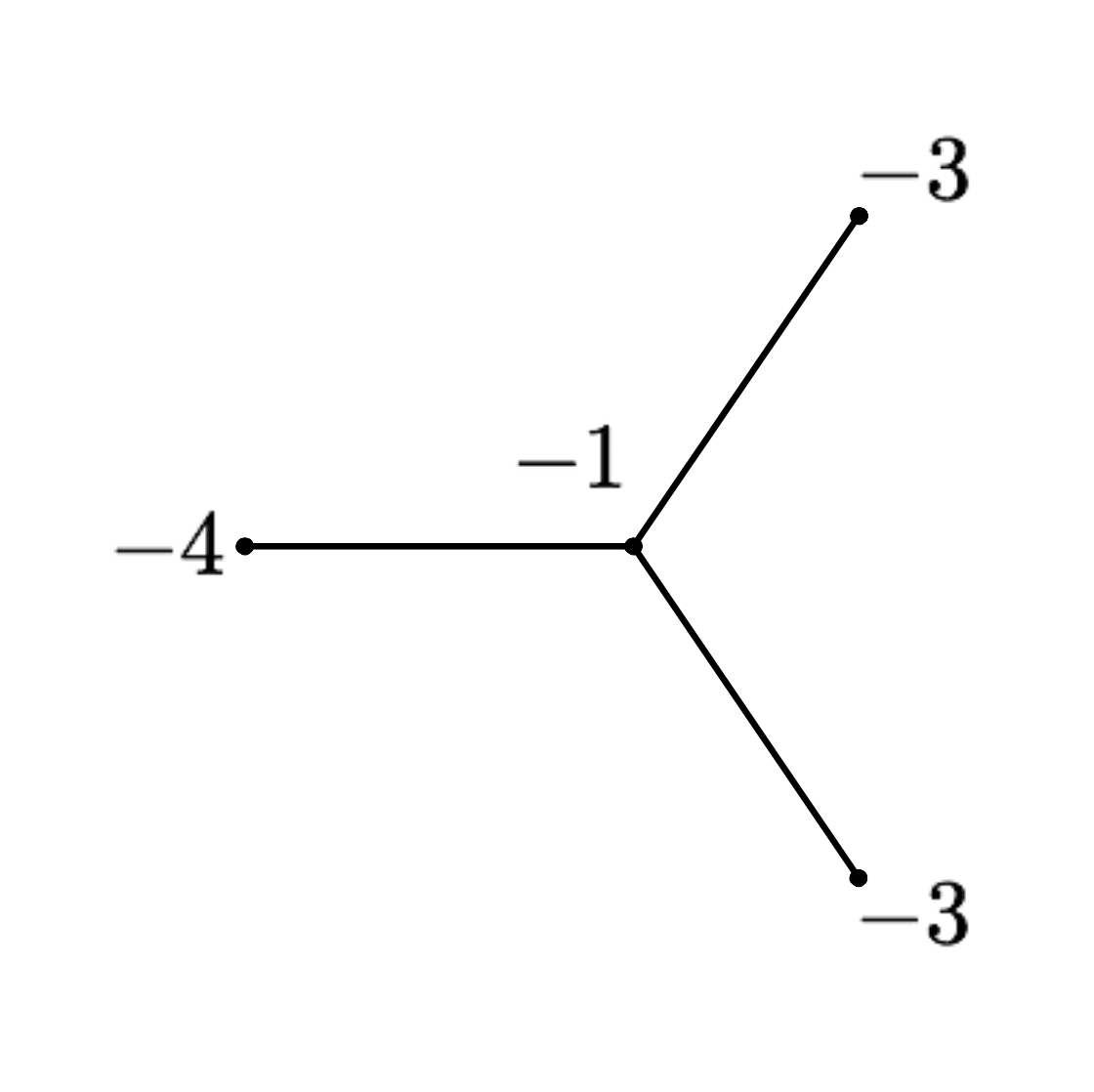}
    \caption{Plumbing graph with a degree three vertex.}
    \label{3ptegdiag}
    \end{subfigure}
    \begin{subfigure}{0.45\textwidth}
    \centering
    \includegraphics[scale=0.23]{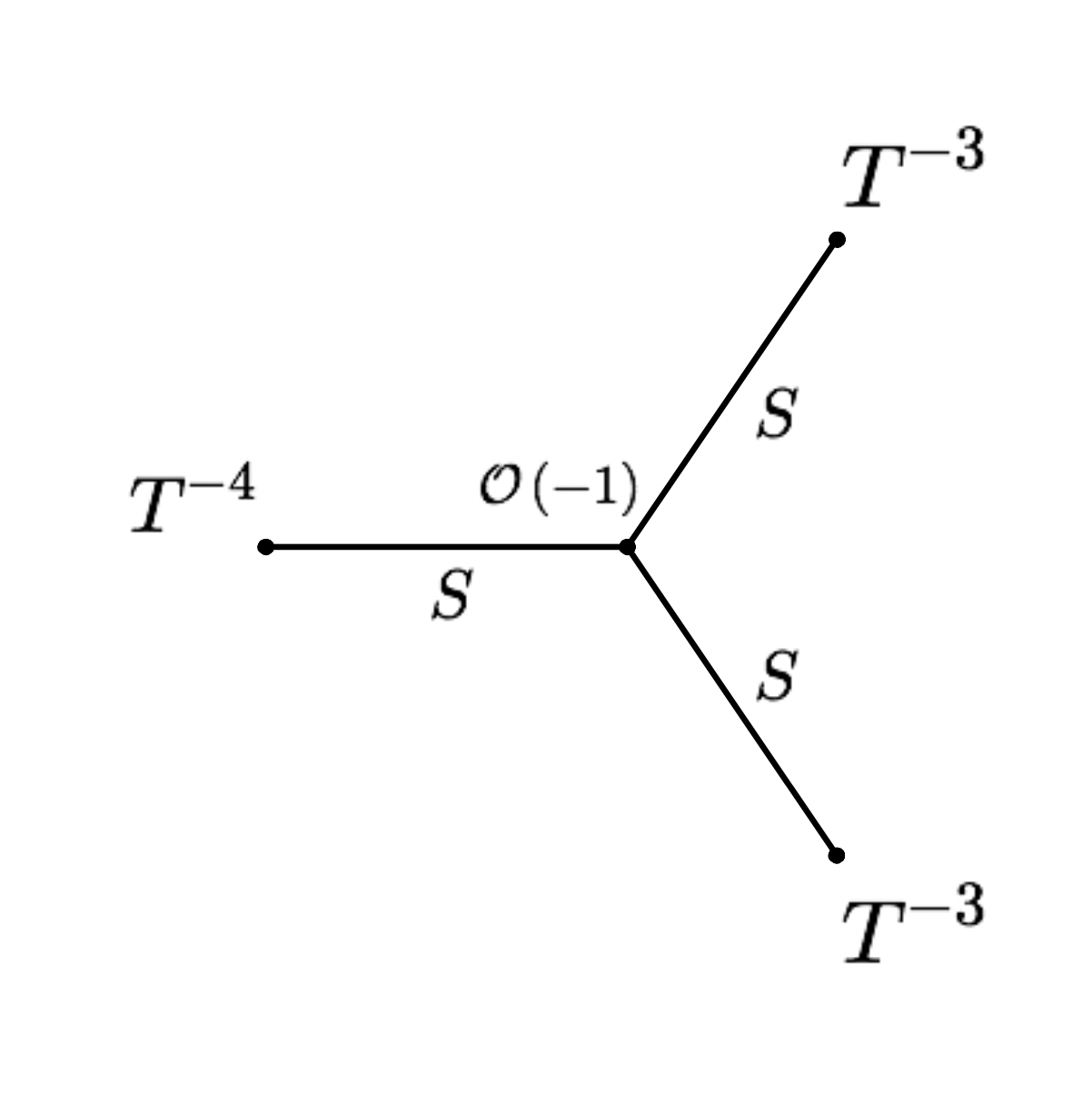}
    \caption{Cutting and gluing of states and operators.}
    \label{3ptegctgl}
    \end{subfigure}
\end{figure}
We can express $\hat{Z}^{\Q/\Z}_{\lambda,\mu}$ as cutting and gluing of states and operators as shown in figure \ref{3ptegctgl}.
\be 
\hat{Z}^{\Q/\Z}_{\lambda,\mu} = q^{-\frac{1}{4}} (v^{\dag} T^{-4} S) \mathcal{O}(-1) (ST^{-3} v)   (D(\lambda,\mu) ST^{-3} v)
\ee 
Where $q^{-\frac{1}{4}}$ is the anomaly contribution. Using the expressions for $v$, $v^{\dag}$, $\mathcal{O}(-1)$, $S$, and $T$ we can compute $\hat{Z}$ and get, 
\begin{align}
    \hat{Z}_{0} &= 1-q+q^{6}-q^{11}+q^{13}-q^{20}+q^{35}-q^{46}+q^{50}-q^{63}+q^{88}+ \cdots   \endline
    \hat{Z}_{\pm 1} &= q^{\frac{2}{3}}(-q+ q^{4}-q^{22}+q^{31}-q^{67}+q^{82}+\cdots).
\end{align}
\end{example}

\subsection*{Bockstein Homomorphism of decorated TQFTs} 

 The decorations of $\hat{Z}$ are the same as those of inverse Reidemeister-Milnor-Turaev torsion. Therefore, as in the case of inverse Reidemeister-Milnor-Turaev torsion, we expect that the Hilbert space that $\hat{Z}$-TQFT assigns to a torus to be decorated by $H^{2}(T^{2},\Z)\cong H_{0}(T^{2},\Z)\cong \Z$, and graded by $H^{1}(T^{2},\Z) \cong \Z^{2}$ (or it's Pontryagin dual with graded trace). Since the surgery formula only computes $\hat{Z}$ for decorations $(0,b)\in \Z^{b_{1}(M_{3})} \times \mathrm{Tors}H_{1}(M_{3})$ we don't expect to see the $H_{0}(T^{2},\Z)\cong \Z$ decoration in $\mathcal{H}(T^{2})$ from equation \eqref{hilbertspaceplumbed}. On the other hand, we do expect to see decoration $b\in \mathrm{Tors}H_{1}(M_{3})$, coming from $H^{1}(T^{2},\Z) \cong \Z^{2}$ grading of the Hilbert space. However, as seen from examples \ref{RLeg}, and \ref{3pteg}, the decoration $b\in \mathrm{Tors}H_{1}(M_{3})$ is coming from the $(\Q/\Z)^{2}$ grading of the Hilbert space. How do we understand this discrepancy? We claim that the $\mathcal{H}(T^{2})$ from equation \eqref{hilbertspaceplumbed} is in fact Hilbert space associated to torus in $\hat{Z}^{\Q/\Z}$-TQFT which under ``Bockstein Homomorphism" maps to $\hat{Z}$-TQFT.
 
 Associated with a short exact sequence of abelian groups 
\begin{align}
    0\rightarrow G_{1} \rightarrow G_{2} \rightarrow G_{3} \rightarrow 0 
\end{align}
there is a connecting homomorphism  $Bk : H^{i}(M_{3},G_{3}) \rightarrow H^{i+1}(M_{3},G_{1}) $ called the Bockstein homomorphism. This Bockstein homomorphism induces a map between topological invariants decorated with $H^{i}(M_{3},G_{3})$ and topological invariants decorated with $H^{i+1}(M_{3},G_{1})$. For $\alpha\in  H^{i}(M_{3},G_{3})$, 
\be 
Z^{\prime}_{\alpha }=Z_{Bk(\alpha)}.  
\ee 
In particular, given a topological invariant decorated with $H^{1}(M_{3},\Q/\Z)$ we get a topological invariant decorated with $H^{2}(M_{3},\Z)$, under the ``Bockstein homomorphism" associated with the following short exact sequence of abelian groups,
\begin{align}
    0\rightarrow \Z \rightarrow \Q \rightarrow \Q/\Z \rightarrow 0.
\end{align}

For plumbed manifolds, with plumbing graph $\Gamma$ and linking matrix $Q$, the cohomology group $H^{1}(M_{3},\Q/\Z)$ is given by,
\be 
H^{1}(M_{3},\Q/\Z) \cong (\Q/\Z)^{b_{1}(\Gamma)} \times Q^{-1}\Z^{V}/\Z^{V} . 
\ee
Where $Q^{-1}\Z^{V}/\Z^{V} =\{ \alpha \in (\Q/\Z)^{V}| Q \alpha \in \Z^{V}   \}$. The Bockstein homomorphism $Bk:H^{1}(M_{3},\Q/\Z)\rightarrow H^{2}(M_{3},\Z)$ takes $(\Q/\Z)^{b_{1}(\Gamma)}$ to $0 \in \Z^{b_{1}(\Gamma)}$, and on $Q^{-1}\Z^{V}$ its given as follows,
\begin{align}
     Bk : Q^{-1}\Z^{V}/\Z^{V} & \rightarrow \Z^{V}/Q\Z^{V}  \nonumber\\
    \alpha & \mapsto Q \alpha \nonumber
\end{align}
Notice the image of $Bk$ is precisely the set of decorations we can get from the surgery formula \eqref{plumbing1}. Thus from equation \eqref{plumbing3} we see that under Bockstein homomorphism, $\hat{Z}^{\Q/\Z}(q)$ maps to $\hat{Z}$,
\be 
  \hat{Z}^{\Q/\Z}_{\beta} (q)= \hat{Z}_{Bk(\beta)}(q)
\ee 

For mapping tori $T^{2}\times_{\varphi}S^{1}$, with $\tr{\varphi}\neq 2$, $H^{1}(T^{2}\times_{\varphi}S^{1},\Q/\Z) \cong \Q/\Z \times (\varphi-I)^{-1} \Z^{2}/\Z^{2}$  and $H^{2}(T^{2}\times_{\varphi}S^{1},\Z) \cong \Z \times \Z^{2}/(\varphi-I)\Z^{2}$, and under Bockstein homomorphism,
\begin{align}
Bk( \Q/\Z \times (\varphi-I)^{-1} \Z^{2}/\Z^{2}) = \{0\} \times \Z^{2}/(\varphi-I)\Z^{2} \subset \Z \times \Z^{2}/(\varphi-I)\Z^{2}
\end{align}
Therefore, the Bockstein homomorphism takes the $(\Q/\Z)^{2}$-graded Hilbert space associated to torus in $\hat{Z}^{\Q/\Z}$-TQFT to $0$-decorated sector of $\Z^{2}$-graded Hilbert space associated to torus in $\hat{Z}$-TQFT. Since the Bockstein homomorphism maps decorations to decorations we expect the Hilbert space for each grading and decoration to remain same. This suggests that the $0$-decorated sector of Hilbert space associated to torus in $\hat{Z}$-TQFT is given by, 
\be 
\mathcal{H}^{0}_{\hat{Z}} (T^{2}) = \C[\Z^{2} \times \Z^{2}].
\ee 
Or $\mathcal{H}^{0}_{(\ell,m)} (T^{2}) = \C[\Z \times \Z]$. Where $0\in \Z \cong H^{2}(T^{2},\Z)$, and $(\ell,m)\in \Z^{2}\cong H^{1}(T^{2},\Z)$ represent the decorations, and grading of the Hilbert space respectively. This conjecture is based upon assumption that the Bockstein homomorphism only talks to the $\Z^{2}$ and $(\Q/\Z)^{2}$ grading. However, it is possible that the two $\Z^{2}$s in $\C[\Z^{2} \times \Z^{2}]$ are identified due to some identifications. In that case the Hilbert space would just be $\mathcal{H}^{0}_{\hat{Z}} (T^{2}) = \C[\Z^{2}]$.

\section{Other invariants from \texorpdfstring{$\hat{Z}$}{Z}} \label{ZNrWRTT}
The $q$-series invariant, in various limits is related to other three-manifold invariants. These relations have been studied in various different places in literature \cite{Gukov:2016gkn,Chun:2019mal,Costantino:2021yfd}. In this section we summarise these conjectural relations and make comments on how the Hilbert spaces in the TQFTs that compute them are related.  

$N_{r}$ invariants are three-manifold invariants associated with quantum groups at roots of unity\cite{Costantino2014QuantumIO, Blanchet2014NonSS, Blanchet2014NonST}. They are decorated by $H^{1}(M_{3},\C/2 \Z)$. The relation between $\hat{Z}(q)$ and $N_{r}$ invariants was studied in \cite{Costantino:2021yfd}. To get the $N_{r}$ invariants from $\hat{Z}(q)$ we first take the Fourier transform of decorations and then take the $q \rightarrow e^{\frac{2 \pi i}{r} }$ limit. This map depends on value of $r \mod 4$. We can schematically express it as,
\be 
N_{r}(M_{3},\omega) =  \sum_{b \in H^{2}(M_{3},\Z)} c^{CGP}_{\omega,b} \lim_{q\rightarrow e^{\frac{2 \pi i}{r} }} \hat{Z}_{b}(M_{3},q) 
\ee 
Where $\omega\in H^{1}(M_{3},\C/2\Z)$ (for more details we refer to \cite{Costantino:2021yfd}). For mapping tori and with $r= 1 \mod 4$, 
\begin{align}
    N_{r}(M,\omega) &= \frac{r}{2} \mathcal{T}(M,[\omega]) \frac{e^{-\frac{i \pi }{2}\mu(M,s)}}{|\mathrm{Tor}H^{1}(M,\Z)|} \endline &\hspace{3cm} \sum_{a,f \in \Z^{v}/Q Z^{v}} e^{-i \pi \omega(a)} e^{2 \pi i \ell k(a+f,f)} e^{2 \pi i (\frac{1-r}{4}\ell k(a,a))} \hat{Z}_{0}(q)\vert_{q \rightarrow e^{\frac{2 \pi i }{r}}}.
\end{align} 
Where $\mathcal{T}(M,[\omega])$ is the is a suitable version of the Reidemeister torsion, $\mu(M,s)$ is the  $\mod 4$ reduction of Rokhlin invariant and $s$ is a spin-structre. 

The Hilbert space associated to torus in $N_{r}$-TQFT for non-integral decorations is given by $\C[H_{r}]$, where
\begin{align}
    H_{r} &=  \{-(r-1),-(r-3),\ldots,(r-1)\}  & &\text{ if } r= 1 \mod 2 \endline
     H_{r} &=  \{1,3,\ldots,(r-1)\}  & &\text{ if } r= 2 \mod 4.
\end{align}
In this basis the $S$ and $T$ matrices are given by 
\begin{align}
    S_{\lambda_{1},\mu_{1},\lambda_{2},\mu_{2}}^{k_{1},k_{2}} &= \frac{1}{\sqrt{r}} \xi^{-(k_{1}+\lambda_{1})(k_{2}+\lambda_{2})+ \ldots} \delta_{\lambda_{1},-\mu_{2}}\delta_{\mu_{1},\lambda_{2}}  \\ T_{\lambda_{1},\mu_{1},\lambda_{2},\mu_{2}}^{k_{1},k_{2}} &= \xi^{\frac{1}{2}(k_{1}+\lambda_{1})^{2}+ \ldots} \delta_{k_{1},k_{2}}\delta_{\lambda_{1},\lambda_{2}}\delta_{\mu_{1},\mu_{2}-\lambda_{2}}.
\end{align}
Where $k_{i} \in H_{r}$, $\lambda_{i},\mu_{i}\in \C/2\Z$, $\xi= e^{\frac{i \pi }{r}}$, and ``$\ldots$" are terms that depend only on $r$.

We expect that this Hilbert space can be obtained from Hilbert space associated with torus in $\hat{Z}^{\Q/\Z}$-TQFT or $\hat{Z}$-TQFT.  In the limit $q^{\frac{1}{2}} = \xi = e^{\frac{i \pi}{r}}$, $q^{\alpha \lambda}$ is same as $q^{\alpha (\lambda+ 2 r)}$. Therefore, the Hilbert space $\C[\Q^{2}]$ reduces to $\C[(\Q/2\Z)^{2} \times (2\Z/ 2 r \Z)^{2}]$ or $\C[\Z^{4}]$ reduces to $\C[\Z^{2}\times (2\Z/ 2 r \Z)^{2}]$. Now taking the Pontryagin dual of grading we get $\C[(\C/2\Z)^{2}\times (2\Z/ 2 r \Z)^{2}]$.  We suspect this can be further reduced to the above Hilbert space and that Gauss sums would play an important role in the reduction giving the $r \mod 4$ dependence of the Hilbert space. 
\begin{center}
    \begin{tikzcd}
  \C[\Z^{4}] \ar[r, "\text{Pontryagin dual}"]  & [2.65em]  \C[(\C/2\Z\times\Z)^{2}] \ar[r, "q^{\alpha \lambda} \sim q^{\alpha (\lambda+2r)}"] & [2.35em]  \C[(\Z\times 2 \Z/2 r\Z)^{2} ] \ar[r]  & \C[\Z^{2} \times 2 \Z/2 r\Z ].
\end{tikzcd}
\end{center}

Similarly, appropriately summing over decorations of $\hat{Z}$ and taking the $q\rightarrow e^{\frac{2 \pi i}{k} }$ limit as conjectured in \cite{Gukov:2017kmk,Chun:2019mal,Costantino:2021yfd} we get the $WRT$ invariants. On the Hilbert space side taking the $q\rightarrow e^{\frac{2 \pi i}{k} }$ limit, $ \C[\Z^{4}] $ reduces to $\C[\Z^{2} \times  \Z/ k\Z ]$ and upon summing over decorations it further reduces to $\C[\Z^{2} \times  \Z/ k\Z ]$.

Without taking the Pontryagin dual of decorations or summing over them, but taking the $q \rightarrow 1$ limit of $\hat{Z}_{b}(q)$, we get the inverse Reidemeister-Milnor-Turaev torsion $\tau^{-1}_{b}$. Therefore, the Hilbert space associated torus in $\hat{Z}$-TQFT should roughly be the same as the one in $\tau^{-1}$-TQFT. However, some states might get identified with each other in the $q\rightarrow 1$ limit. For example, the $0$-decorated sector in $\tau^{-1}$-TQFT is given by $\mathcal{H}_{\ell,m}^{0} = \C$. However, as conjectured in the previous section, the $0$-decorated sector of Hilbert space associated to torus in $\hat{Z}$-TQFT is given by $\mathcal{H}_{\ell,m}^{0}= \C[\Z^{2}] $. We suspect that in the $q\rightarrow1$ limit, $\C[\Z^{2}]$ in $\hat{Z}$-TQFT reduces to $\C$ in $\tau^{-1}$-TQFT, just as it reduced to $\C[2\Z/2r\Z]$ in $N_{r}$-TQFT. 

Using this intuition we conjecture that the Hilbert space associated to torus in $\hat{Z}$-TQFT is given by,
\be 
\mathcal{H}_{\hat{Z}} (T^{2}) = \mathcal{H}_{(2,2)} \otimes \C[\Z^{2}\times \Z^{2}]. 
\ee 
The $ H^{1}(T^{2},\Z) \cong \Z$ decoration comes from the particle number grading of $ \mathcal{H}_{(2,2)}$ while the $ H^{2}(T^{2},\Z) \cong \Z^{2}$ grading comes from the $\Z^{2}$ grading of $\C[\Z^{2}\times \Z^{2}]$. Thinking of $\hat{Z}$-TQFT as $SL(2,\C)$ Chern-Simons theory we could interpret the second $\Z^{2}$ as states created by inserting Wilson lines in solid tori, now taking values in all of $\Z$ as the level is not quantized. 

This intuitive understanding of Hilbert space associated to torus leads us to the conjecture that the Hilbert space associated to genus $g$ surface $\Sigma_{g}$ in the $\hat{Z}$-TQFT is given by 
\be 
\mathcal{H}_{\hat{Z}} (\Sigma_{g}) = \mathcal{H}_{(2g,2g)} \otimes \C[\Z^{2g}\times \Z^{2g}].
\ee 
We note that it is possible that the two $\Z^{2g}$s in $\C[\Z^{2g} \times \Z^{2g}]$ are identified due to some identifications. In that case the Hilbert space would just be $\mathcal{H}_{(2g,2g)} \otimes \C[\Z^{2g}]$. We suspect that the recent progress towards finding a fully general mathematical definition of $\hat{Z}$ from the theory of quantum groups \cite{Park:2020edg, Park:2021ufu}, would provide insights into the validity of above conjecture.

\section*{Acknowledgement}
We are grateful to Sergei Gukov for his guidance through the course of this project. We would also like to thank Sunghyuk Park, and Yixin Xu for insightful conversations during various stages of the project. This work is supported by the Walter Burke Institute for Theoretical Physics,  the U.S. Department of Energy, Office of Science, Office of High Energy Physics, under Award No. DE-SC0011632, and the National Science Foundation under Grant No. NSF DMS 1664227.

\appendix
\section{\texorpdfstring{$\mathrm{Spin}$}{Spin} and \texorpdfstring{$\mathrm{Spin}^{c}$}{Spin c} structures}
The group $\mathrm{Spin}(n)$ is the double cover of the special orthogonal group $SO(n)$ given by the following short exact sequence,
\be 
1 \rightarrow \Z_{2} \rightarrow \mathrm{Spin}(n) \rightarrow SO(n) \rightarrow 1.
\ee 
A $\mathrm{Spin}$ structure on an oriented $n$-dimensional manifold is a lift of the structure group of its tangent bundle from $SO(n)$ to $\mathrm{Spin}(n)$. The group $\mathrm{Spin}^{c}(n)$ is defined by the following short exact sequence 
\be 
1 \rightarrow U(1) \rightarrow \mathrm{Spin}^{c}(n) \rightarrow SO(n) \rightarrow 1.
\ee 
Equivalently we can define it as 
\be 
\mathrm{Spin}^{c}(n) = \frac{\mathrm{Spin}(n)\times U(1)}{\Z_{2}}.
\ee 
Where $\Z_{2} \subset \mathrm{Spin}(n)\times U(1)$ is given by $\{(1,1),(-1,-1)\}$.
A $\mathrm{Spin}^{c}$ structure on an oriented $n$-dimensional manifold is a lift of the structure group of its tangent bundle from $SO(n)$ to $\mathrm{Spin}^{c}(n)$.
    
For three-manifolds the space of $\mathrm{Spin}^{c}$ structures on it, $\mathrm{Spin}^{c}(M_{3})$, is a $H^{2}(M_{3})$-torsor. Suppose $M_{3}$ is a three-manifold obtained by integral surgery on a framed oriented link $L$ in $S^{3}$ and suppose $Q$ is a $V\times V$ linking matrix of $L$. Then we can express the cohomology group $H^{2}(M_{3})$ and the set of $\mathrm{Spin}^{c}$ structures on $M_{3}$, $\mathrm{Spin}^{c}(M_{3})$, as follows,
\begin{align}
    H^{2}(M_{3}) & \cong \Z^{V}/ Q \Z^{V}, \\
    \mathrm{Spin}^{c}(M_{3}) & \cong \{ K \in \Z^{V}/2 Q \Z^{V} \hspace{0.1cm} |\hspace{0.1cm} K_{i} = Q_{ii} \mod 2 \}. 
\end{align}

\bibliographystyle{unsrt}
\bibliography{Bibliography}

\end{document}